\documentclass[referee]{aa} 

\usepackage[dvips]{graphicx}
\usepackage{epsfig}
\usepackage{txfonts}

\newcommand{\rot}{\vec{\nabla} \wedge}
\renewcommand{\div}{\mathrm{div}\,}

\graphicspath{{/home/petri/diocotron/} }

\begin{document}

\title{The diocotron instability in a pulsar ``cylindrical''
  electrosphere.}

\author{J\'er\^ome P\'etri \inst{1}}

\offprints{J. P\'etri}

\institute{Max-Planck-Institut f\"ur Kernphysik, Saupfercheckweg 1,
  69117 Heidelberg, Germany.}

\date{Received / Accepted  nov. 2006}


\abstract
{The physics of the pulsar inner magnetosphere remains poorly
  constrained by observations.  Although about 2000~pulsars have been
  discovered to date, little is known about their emission mechanism.
  Large vacuum gaps probably exist and a non-neutral plasma made of
  electrons in some regions and of positrons in some other regions
  fills space to form an electrosphere.}
{The purpose of this work is to study the stability properties of the
  differentially rotating equatorial disk in the pulsar's
  electrosphere for which the magnetic field is assumed to be dipolar.
  In contrast to previous studies, the magnetic field is not
  restricted to be uniform.}
{A pseudo-spectral Galerkin method using Tchebyshev polynomials
  expansion is developed to compute the spectrum of the diocotron
  instability in a non-neutral plasma column confined between two
  cylindrically conducting walls.  Moreover, the inner wall carries a
  given charge per unit length in order to account for the presence of
  a charged neutron star at the centre of the electrosphere.}
{We show several eigenfunctions and eigenspectra obtained for
  different initial density profiles and electromagnetic field
  configurations useful for laboratory plasmas.  The algorithm is very
  efficient in computing the fastest growing modes.  Applications to a
  ``cylindrical'' electrosphere are also shown for several
  differential rotation profiles. It is found that the growth rates of
  the diocotron instability are of the same order of magnitude as the
  rotation rate.}
{The instability develops on a very short timescale and can account
  for very efficient particle diffusion across the magnetic field
  lines as already claimed in a previous work. The exact geometry of
  the confined plasma, let it be a thin disk or a cylinder, does not
  significantly affect the spectrum of the diocotron instability.}
   

\maketitle

\section{INTRODUCTION}

The detailed structure of charge distribution and electric current
circulation in the closed magnetosphere of a pulsar remains poorly
understood. Although it is often assumed that the plasma entirely
fills the space and corotates with the neutron star, it is on the
contrary very likely that it only partly fills it, leaving large
vacuum gaps in between plasma-filled regions. The existence of such
gaps in aligned rotators has been very clearly established by
Krause-Polstorff and Michel (\cite{1985MNRAS.213P..43K},
\cite{1985A&A...144...72K}).  Since then, a number of different
numerical approaches to the problem have confirmed their conclusions,
including some work by Rylov~(\cite{1989Ap&SS.158..297R}),
Shibata~(\cite{1989Ap&SS.161..187S}),
Zachariades~(\cite{1993A&A...268..705Z}),
Neukirch~(\cite{1993A&A...274..319N}), Thielheim and
Wolfsteller~(\cite{1994ApJ...431..718T}), Spitkovsky and
Arons~(\cite{2002ASPC..271...81S}) and by ourselves (P\'etri et al.
\cite{2002A&A...384..414P}).  This conclusion on the existence of
vacuum gaps has been reached from a self consistent solution of
Maxwell's equations in the case of the aligned rotator.  Moreover,
Smith et al.~(\cite{2001MNRAS.322..209S}) have shown by numerical
modeling that an initially filled magnetosphere like the
Goldreich-Julian model evolves by opening up large gaps and stabilises
to the partially filled and partially void solution found by
Krause-Polstorff and Michel~(\cite{1985MNRAS.213P..43K}) and also by
P\'etri et al.~(\cite{2002A&A...384..414P}).  The status of models of
the pulsar magnetospheres, or electrospheres, has been recently
critically reviewed by Michel (\cite{2005RMxAC..23...27M}).  A
solution with vacuum gaps has the peculiar property that those parts
of the magnetosphere which are separated (following magnetic field
lines) from the star's surface by a vacuum region are not corotating
and suffer differential rotation.
  
This rises the question of the stability of such charged plasma flow.
The differential rotation in the equatorial non neutral disk induces
the so-called diocotron and magnetron instabilities, well known to
plasma physicists (O'Neil~\cite{1980PhFl...23.2216O},
Davidson~\cite{Davidson1990}, O'Neil and
Smith~\cite{1992PhFlB...4.2720O}).  In the inner parts of the
magnetosphere, far from the light cylinder, the instability reduces to
its electrostatic form, the diocotron instability. The linear
development of the diocotron instability of a thin differentially
rotating charged disk has been studied by P\'etri et al
(\cite{2002A&A...387..520P}) and shown to proceed at a growth rate
comparable to the star's rotation rate.  The non linear development of
this instability has been studied by P\'etri et al
(\cite{2003A&A...411..203P}), still in the framework of an infinitely
thin disk model. When there is no external source of charges feeding
the magnetosphere, it has been found that the plasma remains confined
in regions close to the equilibrium state as also shown by
Aly~(\cite{2005A&A...434..405A}).  When however the disk can be fed by
some charge source, P\'etri et al (\cite{2003A&A...411..203P}) have
shown that the instability causes a cross-field transport of these
charges in the equatorial disk, evolving into a net out-flowing flux
of charges. Spitkovsky and Arons~(\cite{2002ASPC..271...81S}) have
numerically studied the problem, and concluded that such charge
transport tends to fill the gaps with plasma. Since a filled
magnetosphere is stable to the diocotron instability, it seems
unlikely that the instability could indeed bring the system to a
completely filled state, even though the disk would not remain thin,
as assumed by P\'etri et al.  (\cite{2002A&A...387..520P},
\cite{2003A&A...411..203P}).  It should be noted that the appearance
of a cross-field electric current as a result of the diocotron
instability has been observed by Pasquini and
Fajans~(\cite{2002AIPC..606..453P}) in laboratory experiments in which
charged particles were continuously injected in the plasma column
trapped in a Malmberg-Penning configuration.
    
The aim of this work is to show that the fact that the differentially
rotating charged disk is not infinitely thin does not invalidate our
former conclusion (P\'etri et al.  \cite{2002A&A...384..414P}) that
the growth rate of the instability is as fast as the star's rotation
rate.  The situation opposite to a thin disk is that of an infinitely
thick disk. By this we mean a plasma column, the structure of which is
invariant with the cylindrical coordinate $z$ and in which the
magnetic field is dependent on the cylindrical coordinate $r$ and
oriented parallel to the z direction. Whereas the thin disk model can
conveniently describe perturbations the horizontal size of which is
much larger than the disk's thickness, the (infinitely) thick model
would be more appropriate to perturbations the horizontal size of
which is much less than the disk's thickness.  In the electrostatic
approximation the magnetic field is not altered by the perturbations
and remains straight. From the inertia-less approximation, the charged
fluid velocity is given by eq.  (\ref{eq:Vit}) which implies that
$\vec{E} + \vec{v} \times \vec{B} =0$ and then that $\vec{E} \cdot
\vec{B} =0$. As a result the perturbed electric field and flow are
2-dimensional, since the electric potential perturbation is
independent of z.  It should be kept in mind that the infinitely thick
disk model with a z-aligned magnetic field is in fact meant to
represent a disk of a finite thickness embedded in a 2D potential
magnetic field.  Such 2D fields, like
\begin{equation}
  \label{eq:B2d}
  \vec{B}= B_\mathrm{r}(r,z) \vec{e}_\mathrm{r} + B_\mathrm{z}(r,z) \vec{e}_\mathrm{z}   
\end{equation}
as opposed to 1D fields like $\vec{B} = B_\mathrm{z}(r) \vec{e}_\mathrm{z}$, may be
current free and still have a radial dependence of $B_\mathrm{z}(r,0)$ in the
equatorial plane. In the 1D model, the disk, though not infinitely
thin, is regarded as not being thick enough for the radial field
component $B_\mathrm{r}(r,z)$ to play a role. The component $B_\mathrm{z}(r,0)$ is taken
as an approximation to $ B_\mathrm{z}(r,z)$.  The domain of validity of a 1D
model then is that $B_\mathrm{r}(r,z)$ remains small in the disk, while the
horizontal size of the perturbations is less than the disk's
thickness.  The real situation is in between the infinitely thin and
the infinitely thick disk model. Dealing with a finite thickness disk
would involve the consideration of the 2D magnetic field structure and
the numerical calculation of the solution to the Poisson's equation in
the appropriate geometry.  Obtaining this solution has been the most
time consuming part of the calculations in P\'etri et al.
(\cite{2003A&A...411..203P}) and things would be even worse if a 2D
geometry with a thick disk were to be considered.  Adopting a 1D
unperturbed geometry model considerably alleviates such difficulties
and seems appropriate to reach a conclusion on the range in which the
growth rates of the diocotron instability are to be found when the
infinitely thin disk approximation is relaxed.
  
In this paper we present a numerical analysis of the linear growth of
the diocotron instability in the 1D model. We plan to compute the full
non linear development of the instability in the same geometry, by
using a 2D electrostatic PIC code, a study that would also be relevant
to laboratory setups. 

The diocotron spectrum for cylindrical geometry and uniform external
magnetic field has been investigated analytically by
Levy~(\cite{1965Levy}) and numerically by Goswami et
al.~(\cite{1999PhPl....6.3442G}) who used a finite difference schemes.
Variational techniques for toroidal non-neutral plasmas have also been
used, see Bhattacharyya~(\cite{2000PhPl....7.4805B}).

The paper is organized as follows.  In Sec.~\ref{sec:Setup}, we
describe the initial setup of the plasma column in the laboratory
consisting of an axially symmetric equilibrium between two conducting
walls.  The slightly different approach used for the pulsar's
electrosphere is also described. In Sec.~\ref{sec:AnalLin}, we recall
the generalised linear eigenvalue problem satisfied by the perturbed
electric potential.  Next, in Sec.~\ref{sec:Numerics}, the
pseudo-spectral Galerkin numerical algorithm to compute the diocotron
spectrum is presented.  Finally, some typical spectra for laboratory
plasma are shown in Sec.~\ref{sec:Results}. We discuss in more detail
the application to electrospheric plasmas which is the main aim of
this work. The conclusions and the possible generalisation are
presented in Sec.~\ref{sec:Conclusion}.

\section{INITIAL SETUP}
\label{sec:Setup}

In laboratory plasmas, the motion of particles is imposed by external
devices controlling the potential and electric field.  It is the most
common situation. However, having in mind to applied the algorithm to
pulsar's electrosphere, it is also interesting to study the stability
properties of non-neutral plasmas by imposing the rotation profile
instead of the density profile. Therefore, we first describe the case
encountered in the laboratory and then discussed how to apply and
extend it to pulsars.

\subsection{Laboratory plasma}

We consider a one species non-neutral plasma consisting of particles
with mass~$m$ and charge~$q$ trapped between two cylindrically
conducting walls located at $r = W_1$ and $r = W_2 > W_1$. The plasma
column itself is confined between $R_1 \ge W_1$ and $R_2 \le W_2$.
This allows us to take into account vacuum regions between the plasma
and the conducting walls. We adopt cylindrical coordinates denoted
by~$(r,\varphi,z)$ and the corresponding basis
vectors~$(\vec{e}_\mathrm{r},\vec{e}_\varphi,\vec{e}_\mathrm{z})$.  In order to simulate
the presence of a charged neutron star generating a radial electric
field, the inner wall at~$W_1$ carries a charge~$Q$ per unit length
such that its electric field is simply given by Maxwell-Gauss
theorem~:
\begin{equation}
  \label{eq:E_r_ext}
  \vec{E}_\mathrm{w}(r) = \frac{Q}{2\,\pi\,\varepsilon_0\,r} \, \vec{e}_\mathrm{r}
\end{equation}
In the equilibrium configuration, the particle number density
is~$n(r)$ and the charge density is~$\rho(r) = q \, n(r)$. In contrast
to earlier studies, the external magnetic field, along the~$z$-axis,
is not necessarily uniform
\begin{equation}
  \label{eq:Bz}
  \vec{B} = B_\mathrm{z}(r) \, \vec{e}_\mathrm{z}
\end{equation}
Strictly speaking, this magnetic field is associated with an azimuthal
current because $\rot \vec{B} \ne \vec{0}$. However, as claimed in the
introduction, there is also a (weak) radial component
Eq.~(\ref{eq:B2d}) which should compensate for this current. Let's
find the conditions for which approximation Eq.~(\ref{eq:Bz}) holds,
assuming a power law for $B_\mathrm{z}$, Eq.~(\ref{eq:BF}).  From
Eq.~(\ref{eq:B2d}), the $\varphi$-component of the curl reads
\begin{equation}
  \label{eq:cdurl}
  \frac{\partial B_\mathrm{r}}{\partial z} - \frac{\partial B_\mathrm{z}}{\partial r}
  = \frac{\partial B_\mathrm{r}}{\partial z} + \alpha \frac{B_\mathrm{z}}{r}
\end{equation}
With the boundary conditions $B_\mathrm{r}(r,z=0)=0$, we found that near the
equatorial plane
\begin{equation}
  \label{eq:BR}
  B_\mathrm{r} = - \alpha \, \frac{z}{r} \, B_\mathrm{z}
\end{equation}
$\alpha$ being of order unity, the radial component of the magnetic
field is negligible whenever $z\ll r$. Therefore, our approximation
Eq.~(\ref{eq:Bz}) holds in the thin disk limit (intermediate between
infinitely thin and infinitely thick).

The electric field is made of two parts, the first one arising from
the plasma column~$\vec{E}_\mathrm{p}$ itself, and the second one from
the inner conducting wall~$\vec{E}_\mathrm{w}$,
Eq.~(\ref{eq:E_r_ext}).  We assume that the electric field induced by
the plasma vanishes at $r=W_1$, i.e.  $\vec{E}_\mathrm{p}(W_1) =
\vec{0}$. We therefore get
\begin{equation}
  \label{eq:Ep}
  \vec{E}_\mathrm{p}(r) = \frac{1}{\varepsilon_0\,r} \, \int_{W_1}^r 
  \rho(r') \, r' \, dr' \, \vec{e}_\mathrm{r}
\end{equation}
The total electric field, directed along the radial
direction~$\vec{e}_\mathrm{r}$, is therefore
\begin{equation}
  \label{eq:ETot}
  \vec{E} = \vec{E}_\mathrm{p} + \vec{E}_\mathrm{w} = E_\mathrm{r} \, \vec{e}_\mathrm{r}
\end{equation}
At equilibrium, the plasma is rotating between the two walls at a
speed $\Omega(r)$. The particles have mass~$m_\mathrm{e}$ and charge~$q$. In
the stationary state, the centrifugal force is balanced by the Lorentz
force such that
\begin{equation}
  \label{eq:FL}
  q \, ( E_\mathrm{r} + r \, \Omega \, B_\mathrm{z} ) + m_\mathrm{e} \, r \, \Omega^2 = 0
\end{equation}
Introducing the cyclotron frequency by~$\omega_\mathrm{c} = \frac{q\,B_\mathrm{z}}{m_\mathrm{e}}$
and the plasma frequency by~$\omega_\mathrm{p}^2 = \frac{n\,q^2}
{m_\mathrm{e}\,\varepsilon_0}$, the angular speed of the column satisfies the
quadratic equation
\begin{eqnarray}
  \label{eq:Omega}
  \Omega^2 + \omega_\mathrm{c} \, \Omega + \frac{q\,E_\mathrm{r}}{m_\mathrm{e}\,r} & = & 0 
\end{eqnarray}
leading to two possible solutions for $\Omega$, namely
\begin{eqnarray}
  \label{eq:sol}
  \Omega & = & - \frac{\omega_\mathrm{c}}{2} \, \left[ 1 \pm 
    \sqrt{ 1 - 4 \, \frac{q\,E_\mathrm{r}}{m_\mathrm{e}\,r\,\omega_\mathrm{c}^2}} \right] 
\end{eqnarray}
If the electric field is sufficiently weak, i.e. $q\,E_\mathrm{r} \ll
m_\mathrm{e}\,r\,\omega_\mathrm{c}^2$, the solutions are approximated
by $\Omega = - \omega_\mathrm{c}$ and
\begin{equation}
  \label{eq:DriftElect}
  \Omega = - \frac{E_\mathrm{r}}{r\,B_\mathrm{z}}
\end{equation}
The weak field limit is equivalent to a low density plasma column
because it implies~$\omega_\mathrm{p}^2 \ll \omega_\mathrm{c}^2$.  The solution
Eq.~(\ref{eq:DriftElect}) corresponds to the usual electric drift
approximation (Davidson and Tsang~\cite{1984PhRvA..30..488D}). In the
remaining of this paper, we retain this assumption. Therefore, the
equation of motion for the charged particles is replaced by the
electric drift approximation.  Indeed, we consider only the
low-density non-neutral plasma case for which the diocotron regime is
valid.  Neglecting electron inertia, the electric drift approximation
applies. For $Q=0$ and a constant density profile in the plasma column
with $W_1=0$, it corresponds to a circular motion at the diocotron
frequency defined by, see for instance Davidson~(\cite{Davidson1990})
\begin{equation}
  \label{eq:DioFreq}
  \omega_D = \frac{\omega_\mathrm{p}^2}{2\,\omega_\mathrm{c}}
\end{equation}

\subsection{Electrospheric plasma}

Our purpose is to study the non-neutral plasma instabilities occurring
in the pulsar inner magnetosphere. The most interesting feature
obtained in the work by P\'etri et al.~(\cite{2002A&A...384..414P}) by
constructing an electrospheric model was a differential rotation in
the positively charged equatorial disk. It can indeed be shown that
large vacuum gaps imply a significant departure from corotation. That
is why when applying our algorithm to pulsars, it is more appropriate
to assume a given rotation profile in the plasma~$\Omega(r)$, instead
of densities and potential imposed by external devices. In the
electric drift approximation, using Maxwell-Gauss law, the charge
density is recovered according to ideal MHD by
\begin{equation}
  \vec{E} = - ( \vec{\Omega} \wedge \vec{r} ) \wedge \vec{B}  
\end{equation}
leading to the charge density as follows
\begin{equation}
  \label{eq:DensiteCharge}
  \rho = - \varepsilon_0 \, \left[ \frac{B_\mathrm{z}}{r} \,
    \frac{\partial}{\partial r} \left( r^2 \, \Omega \right)
    + r \, \Omega \, \frac{\partial B_\mathrm{z}}{\partial r} \right]
\end{equation}
Therefore, there are only 2 independent functions, namely the
couple~$(B_\mathrm{z},\Omega)$ appropriate for pulsar electrospheric plasmas and
the couple~$(B_\mathrm{z},\rho)$ useful for laboratory plasmas.

\section{LINEAR ANALYSIS}
\label{sec:AnalLin}

We consider two different plasma configurations. In the first case,
the plasma column is supposed to be in contact with both the inner and
the outer wall. This will be called single domain. In the second case,
vacuum regions exist between the plasma column and the inner and/or
the outer walls. This will be called multi-domain.  We introduced
these two situations because of the slightly different numerical
treatment of the problem. Indeed, vacuum regions are treated
analytically whereas regions filled with plasma are treated
numerically.

\subsection{Single domain}

Let's start with a column density in contact with the inner and the
outer conducting wall. The motion of the column of plasma is governed
by the conservation of charge, the Maxwell-Poisson equation and the
electric drift approximation, respectively~:
\begin{eqnarray}
  \label{eq:EqnGMot}
  \frac{\partial\rho}{\partial t} + \div (\rho \, \vec{v}) & = & 0 \\
  \Delta \phi + \frac{\rho}{\varepsilon_0} & = & 0 \\
  \label{eq:Vit}
  \vec{v} & = & \frac{\vec{E} \wedge \vec{B}}{B^2} \\
  \vec{E} & = & -\vec{\nabla} \phi
\end{eqnarray}
We recall that the magnetic field remains constant in time in the
diocotron regime. We apply the standard linear perturbation theory.
Introducing perturbations of physical quantities~$X$ like electric
potential, density and velocity components, by the expansion
\begin{equation}
  \label{eq:Expa}
  X(r,\varphi,t) = X(r) \, e^{i\,(m\,\varphi - \omega\,t)}
\end{equation}
the eigenvalue problem for the perturbed electric potential~$\phi(r)$
is expressed as
\begin{equation}
  \label{eq:SYsvl}
  \left[ \frac{1}{r} \, \frac{\partial}{\partial r} 
    \left( r \, \frac{\partial\phi}{\partial r} \right) 
    - \frac{m^2}{r^2} \right] \, \phi = 
  \frac{m}{(\omega - m \, \Omega)} \, 
  \frac{1}{\varepsilon_0 \, r} \, \frac{\partial}{\partial r}
  \left( \frac{\rho}{B_\mathrm{z}} \right) \, \phi
\end{equation}
For the purpose of our numerical algorithm, it is more convenient to
rewrite it as a generalised linear eigenvalue problem as follows
\begin{equation}
  \label{eq:Eqw}
  \omega \, \mathcal{L}_m(\phi) = m \, \Omega \, \mathcal{L}_m(\phi) + q_m \, \phi
\end{equation}
The Laplacian operator~$\mathcal{L}_m$ for each azimuthal mode~$m$ is
given by
\begin{equation}
  \label{eq:OpLapl}
  \mathcal{L}_m(\phi) = \left[ \frac{1}{r} \, \frac{\partial}{\partial r} 
    \left( r \, \frac{\partial\phi}{\partial r} \right) 
    - \frac{m^2}{r^2} \right] \, \phi
\end{equation}
and the function~$q_m$ is
\begin{equation}
  q_m = \frac{m}{\varepsilon_0 \, r} \, \frac{\partial}{\partial r} \left(
    \frac{\rho}{B_\mathrm{z}} \right)
\end{equation}
For the electrospheric plasma, we can eliminate $\rho$ by
Eq.~(\ref{eq:DensiteCharge}) and write
\begin{equation}
  q_m = - \frac{m}{r} \, \frac{\partial}{\partial r} \left[
    \frac{1}{r} \frac{\partial}{\partial r} \left( r^2 \, \Omega \right) + 
    \frac{r\,\Omega}{B_\mathrm{z}} \, \frac{\partial B_\mathrm{z}}{\partial r}  \right] 
\end{equation}
Note that this function does not depend on the intensity of~$B_\mathrm{z}$ but
only on its radial profile. However, it depends linearly on the
amplitude of the differential rotation, meaning that strong departure
from corotation leads to strong instability as will be shown in the
next section. The eigenvalue problem Eq.~(\ref{eq:Eqw}) is
supplemented by homogeneous boundary conditions such that the
perturbed potential vanishes at the boundaries, namely
\begin{equation}
  \label{eq:Dirichlet}
  \phi(W_1) = \phi(W_2) = 0
\end{equation}

\subsection{Multi-domain decomposition}

When vacuum gaps exist between the walls and the plasma, the
discontinuity in the density profile introduces a Gibbs phenomenon and
therefore drastically decreases the efficiency of our pseudo-spectral
algorithm.  To overcome this difficulty, we decompose the space
between the two walls into three distinct regions, namely
\begin{itemize}
\item region~I: vacuum space between inner wall and inner boundary of
  the plasma column, the solution for the electric potential is
  denoted by $\phi_\mathrm{I}$, defined for $W_1 \le r \le R_1$~;
\item region~II: the plasma column itself located between $R_1$ and
  $R_2$, solution denoted by $\phi_\mathrm{II}$, defined for $R_1
  \le r \le R_2$~;
\item region~III: vacuum space between outer boundary of the plasma
  column and the outer wall, solution denoted by $\phi_\mathrm{III}$,
  defined for $R_2 \le r \le W_2$.
\end{itemize}
In region I and III, the vacuum solutions which satisfy the required
boundary conditions ($\phi_\mathrm{I}(W_1)=0$ and
$\phi_\mathrm{III}(W_2)=0$) are given by the usual solutions to
Laplace equation in the form
\begin{eqnarray}
  \label{eq:Phi1et3}
  \phi_\mathrm{I}  (r) & = & A \, \left( r^m - \frac{W_1^{2m+1}}{r^{m+1}} \right) \\
  \phi_\mathrm{III}(r) & = & B \, \left( r^m - \frac{W_2^{2m+1}}{r^{m+1}} \right)
\end{eqnarray}
where $A$ and $B$ are two constants to be determined by matching the
boundary conditions at the interfaces between plasma and vacuum.
Because no surface charges accumulate on these interfaces (the
equilibrium density profile should vanish at $R_1$ and $R_2$), the
electric field is continuous in the whole space. The matching
conditions are therefore, continuity of $\phi$ and of its first
derivatives, namely
\begin{eqnarray}
  \label{eq:Matching}
  \phi_\mathrm{I}  (R_1) & = & \phi_\mathrm{II}  (R_1) \\
  \phi_\mathrm{I}' (R_1) & = & \phi_\mathrm{II}' (R_1) \\
  \phi_\mathrm{III} (R_2) & = & \phi_\mathrm{II}  (R_2) \\
  \phi_\mathrm{III}'(R_2) & = & \phi_\mathrm{II}' (R_2)
\end{eqnarray}
We eliminate $A$ and $B$ from these equations to get the boundary
conditions for~$\phi_\mathrm{II}$.  We find that in the plasma column,
the potential has to satisfy mixed boundary conditions of Robin type
\begin{equation}
  \label{eq:Robin}
  \alpha_{1/2} \, \phi_\mathrm{II}(R_{1/2}) + \beta_{1/2} \, \phi_\mathrm{II}'(R_{1/2}) = 0
\end{equation}
with the coefficients given by
\begin{eqnarray}
  \label{eq:BLRobin}
  \alpha_{1/2} & = & \left( m \, R_{1/2}^{m-1} + 
    ( m + 1 ) \, \frac{W_{1/2}^{2m+1}}{R_{1/2}^{m+2}} \right)  \\
  \beta_{1/2} & = & - \, \left( R_{1/2}^m - 
    \frac{W_{1/2}^{2m+1}}{R_{1/2}^{m+1}} \right)
\end{eqnarray}
The boundary conditions Eq.~(\ref{eq:Robin}) are the generalisation of
the homogeneous Dirichlet case Eq.~(\ref{eq:Dirichlet}) when vacuum
gaps are present. If the vacuum regions were removed such that
$W_1=R_1$ and $W_2=R_2$, the coefficients $\beta_{1/2}$ in
Eq.~(\ref{eq:Robin}) would vanish whereas $\alpha_{1/2}\ne0$. We
therefore find again the homogeneous boundary conditions of the
previous subsection, Eq.~(\ref{eq:Dirichlet}).

\section{NUMERICAL PROCEDURE}
\label{sec:Numerics}

\subsection{Single domain}

In the single domain decomposition, the plasma is in contact with
both, the inner and outer wall. Thus $W_1=R_1$ and $W_2=R_2$. There is
no need to distinguish between $R$ and $W$. The generalised linear
eigenvalue problem Eq.~(\ref{eq:Eqw}) is solved numerically by using a
pseudo-spectral Galerkin method taking advantage of the homogeneous
boundary conditions imposed on the perturbed electric potential,
$\phi(R_{1/2}) = 0$. First, the independent variable~$r$ in
Eq.~(\ref{eq:Eqw}) is transformed into a new independent variable~$x$
such that~$r\in[R_1,R_2]$ is mapped into the interval~$x\in[-1,1]$.
This coordinate transformation reads~:
\begin{equation}
  \label{eq:TransfoXR}
  r = \frac{R_2-R_1}{2} \, x + \frac{R_2+R_1}{2}
\end{equation}
Second, the unknown eigenfunction~$\phi$ is expanded into a set of
basis functions~$(\psi_k)_{k\ge2}$ (Boyd~\cite{Boyd2001}) defined for
$n\ge1$ by~:
\begin{eqnarray}
  \label{eq:Base1}
  \psi_{2n  }(x) & \equiv & T_{2n  }(x) - 1 \\
  \label{eq:Base2}
  \psi_{2n+1}(x) & \equiv & T_{2n+1}(x) - x
\end{eqnarray}
Therefore the boundary conditions are automatically satisfied for each
function of this particular basis, $\psi_k(\pm1)|_{k\ge2} = 0$.  Here
$T_k(x) = \cos(k \, \arccos(x))$ are the Tchebyshev polynomials. Let
$N$ being the number of collocation points $(x_i)_{0\le i\le N-1}$ for
the Tchebyshev expansion~:
\begin{equation}
  \label{eq:Coloc}
  x_i = \cos\left( \frac{i \, \pi}{N-1} \right) 
  \;\;\;\;\;\;\;\;\;\;\;\;   i=0,1,..,N-1
\end{equation}
The unknown function is expanded according to~:
\begin{equation}
  \label{eq:Dvlpt}
  \phi(x) = \sum_{k=2}^{N-1} \, \phi_k \, \psi_k(x)
\end{equation}
The matrix discretisation of Eq.~(\ref{eq:Eqw}) leads to the
generalised linear eigenvalue problem
\begin{equation}
  \label{eq:SystGeneral}
  L \, \vec{\phi} = \omega \, M \, \vec{\phi}
\end{equation}
where we have introduced the unknown vector $\vec{\phi} = (\phi_2,
\phi_3, ..., \phi_{N-1})$ and the square matrices are defined by~:
\begin{eqnarray}
  \label{eq:MatrixLM}
  L_{i-1,k-2} & \equiv & m \, \Omega(x_i) \, \mathcal{L}_m(\psi_k)(x_i) +
  q_m(x_i) \, \psi_k(x_i) \\
  M_{i-1,k-2} & \equiv & \mathcal{L}_m(\psi_k)(x_i) \\
  i & = & 1,2,..,N-2 \nonumber \\
  k & = & 2,3,...,N-1 \nonumber 
\end{eqnarray}
The system Eq.~(\ref{eq:SystGeneral}) is then solved by standard
routines to compute the eigenvalues and the eigenvectors.

\subsection{Multi-domain decomposition}

In the multi-domain decomposition, the Galerkin basis
Eq.~(\ref{eq:Base1})-(\ref{eq:Base2}) is not adapted anymore. The new
basis functions have to satisfy the following Robin boundary
conditions~:
\begin{eqnarray}
  \label{eq:RobinBase1}
  \alpha_1 \, \psi_k(-1) + \beta_1 \, \psi_k'(-1) & = & 0 \\
  \label{eq:RobinBase2}
  \alpha_2 \, \psi_k(+1) + \beta_2 \, \psi_k'(+1) & = & 0
\end{eqnarray}
We look for basis functions expressed by a three term expansion in
Tchebyshev polynomials, for $k\ge2$, like
\begin{equation}
  \label{eq:Base3}
  \psi_k(x) = T_k(x) + a_k \, T_{k-1}(x) + b_k \, T_{k-2}(x)
\end{equation}
For $\psi_k(x)$ to satisfy the conditions
Eq.~(\ref{eq:RobinBase1})-(\ref{eq:RobinBase2}), the unknowns~$a_k$
and~$b_k$ have to be solutions of the following system
\begin{eqnarray}
  \label{eq:SystBase}
  a_k \, \left[ \beta_1 \, (k-1)^2 - \alpha_1 \right] + 
  b_k \, \left[ \alpha_1 - \beta_1 \, (k-2)^2 \right] & = & - \alpha_1 + \beta_1 \, k^2  \\
  a_k \, \left[ \beta_2 \, (k-1)^2 + \alpha_2 \right] + 
  b_k \, \left[ \alpha_2 + \beta_2 \, (k-2)^2 \right] & = & - \alpha_2 - \beta_2 \, k^2 
\end{eqnarray}
The procedure to discretise the eigenvalue problem is the same as the
one described in the previous section. We need only to replace the old
basis Eq.~(\ref{eq:Base1})-(\ref{eq:Base2}) by the new basis
Eq.~(\ref{eq:Base3}).

\section{RESULTS}
\label{sec:Results}

Using the aforementioned numerical algorithm, we compute the
eigenfunctions and eigenspectra of the diocotron instability for
various equilibrium density profiles and charges on the inner wall for
laboratory plasmas, Sec.~\ref{subsec:Lab} and several rotation
profiles for pulsar's electrosphere, Sec.~\ref{subsec:Pulsar}.

In all computations, the external magnetic field follows a decreasing
power law with index~$\alpha \ge 0$ such that
\begin{equation}
  \label{eq:BF}
  B_\mathrm{z} = \frac{B_0}{r^\alpha} 
\end{equation}

\subsection{Plasma column}
\label{subsec:Lab}

First, we consider the laboratory plasma confined by some external
experimental electromagnetic devices. The magnetic field and the
density profile are specified as initial data. The rotation profile is
then deduced from the electric drift approximation Eq.~(\ref{eq:Vit}).

\subsubsection{Constant density profile}

The simplest case corresponds to a uniform magnetic field~$B_0$,
$\alpha = 0$ and a constant charge density~$\rho_0$ extending from the
origin~$r=0$ to the outer wall, so~$W_1=R_1=0$, no inner conducting
wall, $W_2=R_2$ and also $Q=0$ (single domain).  In this particular
case, the equilibrium speed is constant~:
\begin{equation}
  \label{eq:Const}
  \Omega = - \frac{\rho_0}{2\,\varepsilon_0\,B_0}
\end{equation}
The functions~$q_m$ are all identically zero and the analytical
solution to the eigenvalue problem Eq.~(\ref{eq:SystGeneral}) is
simply~:
\begin{equation}
  \label{eq:solanalyt}
  \omega = m \, \Omega
\end{equation}
There is no instability.  Our numerical algorithm is able to reproduce
the eigenvalues Eq.~(\ref{eq:solanalyt}) with great accuracy taking
only a few discretisation points (9~points are already enough). The
convergence is very fast due to the evanescent truncation error
introduced by the Tchebyshev expansion. The precision achieves easily
10~digits.

\subsubsection{Decreasing density profile}

\begin{figure}[htbp]
  \centering
  \includegraphics[scale=0.8]{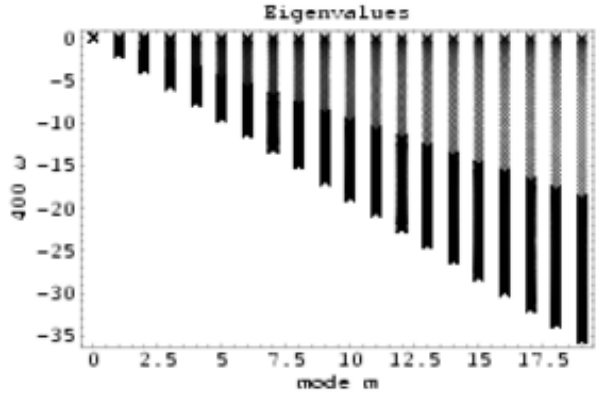}
  \caption{Eigenvalues for the monotonely decreasing 
    density profile Eq.~(\ref{eq:Momto}). The boundaries are set to
    $W_1 = R_1 = 1$ and $W_2 = R_2 = 20$. The other parameters are
    $\alpha=0$ and $Q=0$.  The continuous spectrum is shown for the
    azimuthal modes $m=0..19$. The eigenfrequencies have been
    normalised to the rotation frequency $\omega_\mathrm{r} = \omega_\mathrm{p}^2 / 4 \,
    \omega_\mathrm{c} = 1/400$.}
  \label{fig:DDP1}
\end{figure}

\begin{figure}[htbp]
  \centering \includegraphics[scale=0.8]{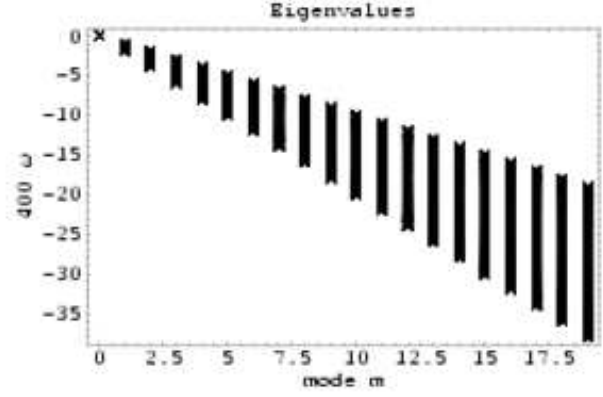}
  \caption{Same as Fig.~\ref{fig:DDP1} but the inner wall 
    has been removed, i.e. $W_1 = R_1 = 0$.}
  \label{fig:DDP2}
\end{figure}

\begin{figure}[htbp]
  \centering
  \includegraphics[scale=0.8]{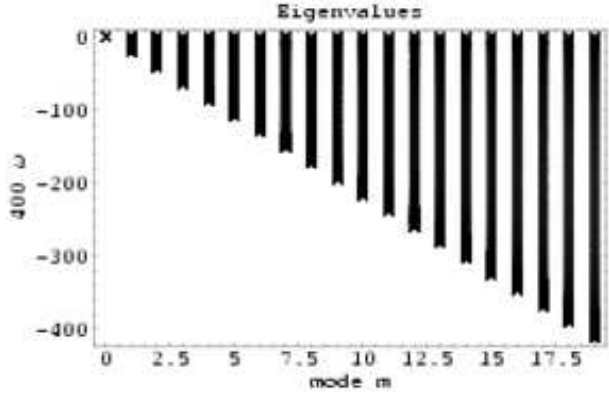}
  \caption{Same as Fig.~\ref{fig:DDP1} but with a decreasing
    external magnetic field, $\alpha=1$. Note the difference in scale
    for both cases. }
  \label{fig:DDP3}
\end{figure}

Next we consider a monotonely decreasing density profile starting from
$\rho_0$ at $r=W_1=R_1$ and vanishing at $r=W_2=R_2$ (single domain)
such that
\begin{equation}
  \label{eq:Momto}
  \rho = \rho_0 \, \frac{R_2^2 - r ^2}{R_2^2 - R_1^2}
\end{equation}
The corresponding rotation profile can be found analytically. The
electric field induced by the plasma is deduced from Eq.~(\ref{eq:Ep})
whereas the contribution from the wall is given by
Eq.~(\ref{eq:E_r_ext}). Using the drift approximation
Eq.~(\ref{eq:DriftElect}), the total radial electric field expressed
in Eq.~(\ref{eq:ETot}) and the power law index for the magnetic field
Eq.~(\ref{eq:BF}), the rotation speed of the plasma column in the most
general case ($\alpha$ and $Q$ arbitrary) is equal to
\begin{equation}
  \label{eq:se}
  \Omega = - \frac{\rho_0 \, r^\alpha}
  {\varepsilon_0 \, B_0 \, ( R_2^2 - R_1^2 )}
  \left[ \frac{R_2^2}{2} \, \left( 1 - \frac{R_1^2}{r^2} \right)
    - \frac{1}{4} \, \left( r^2 - \frac{R_1^4}{r^2} \right) \right] -
  \frac{Q}{2\,\pi\,\varepsilon_0\,r^{2-\alpha} \, B_0}
\end{equation}
When the plasma is in contact with the inner as well as with the outer
conducting wall and the applied magnetic field is uniform, there is no
diocotron instability, Briggs et al.~(\cite{Briggs}). Indeed, for
$\alpha=0$, all the eigenvalues we found are real as expected. For the
first azimuthal modes~$m\le19$, the map of the eigenvalues in the
complex plane is shown in Fig.~\ref{fig:DDP1} for an uniform magnetic
field, $\alpha=0$, zero charge, $Q=0$, $W_1=R_1=1$ and $W_2=R_2=20$.
We normalised the frequencies to the rotation frequency $\omega_\mathrm{r} =
\omega_\mathrm{p}^2 / 4 \, \omega_\mathrm{c} = 1/400$. Therefore, for each mode, the
spectrum is continuous and delimited by two integer values, namely
\begin{equation}
  0 \ge \omega_m / \omega_\mathrm{r} \ge -2\,m
\end{equation}
A similar conclusion applies also when the inner wall is removed ($W_1
= R_1 = 0$), see Fig.~\ref{fig:DDP2}. In this case the continuous
spectra satisfy
\begin{equation}
  -m \ge \omega_m / \omega_\mathrm{r} \ge -2\,m
\end{equation}
Finally when switching from an uniform magnetic field to a decreasing
power law~Eq.~(\ref{eq:BF}) with for instance $\alpha=1$, the results
are qualitatively unchanged, see Fig.~\ref{fig:DDP3}.  As can be seen
by comparing the two maps on Fig.~\ref{fig:DDP1} and
Fig.~\ref{fig:DDP3}, the radial profile of the magnetic field does not
influence much the eigenvalues. There is mainly a difference in scale
because the rotation profile for $\alpha=1$ leads to higher rotation
rate due to the extra factor $r^\alpha$ in Eq.~(\ref{eq:se}).

\subsubsection{Gaussian density profile}

Next we consider a situation in which the plasma is not in contact
with neither the inner nor the outer wall and so $W_1<R_1$ and
$W_2>R_2$, (multi-domain decomposition).  We choose a Gaussian density
profile given by
\begin{equation}
  \label{eq:Gaussien}
  \rho = \rho_0 \, e^{- ( r - a )^2 / 2 \, \sigma^2}
\end{equation}
$\sigma$ is a measure of the profile radial spread while $a$ specifies
the location of the centre of the plasma column. The associated
rotation rate of the plasma column in the most general case ($\alpha$
and $Q$ arbitrary) is found analytically to be~:
\begin{eqnarray}
  \label{eq:ses}
  \Omega & = & - \frac{\rho_0 \, \sigma}{\varepsilon_0 \, B_0 \, r^{2-\alpha}}
  \left[ \sigma \, \left\{ e^{ - ( R_1 - a )^2 / 2\sigma^2} -
      e^{ - ( r - a )^2 / 2\sigma^2} \right\} \right. \nonumber \\  
  & + & \left. a \, \sqrt{\frac{\pi}{2}}
    \left\{ \mathrm{erf} \left( \frac{r-a}{\sigma\,\sqrt{2}} \right) -
      \mathrm{erf} \left( \frac{R_1-a}{\sigma\,\sqrt{2}} \right) \right\} \right] -
  \frac{Q}{2\,\pi\,\varepsilon_0\,r^{2-\alpha} \, B_0}
\end{eqnarray}
$\mathrm{erf}(x)$ is the error function, Abramowitz and
Stegun~(\cite{1965hmfw.book.....A}).  In this case, we expect a strong
diocotron instability with growth rates of the same order of magnitude
as the rotation speed of the perturbation.

\begin{figure}[htbp]
  \centering
  \includegraphics[scale=0.8]{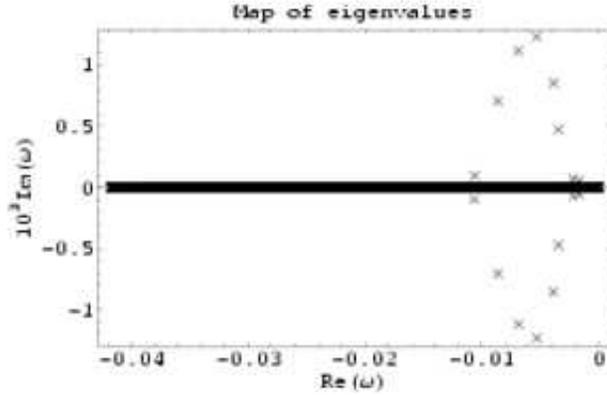}
  \caption{Eigenvalues for the Gaussian density profile Eq.~(\ref{eq:Gaussien})
    with $\sigma=1$ and $a=7$.  The boundaries are set to $W_1=1$,
    $W_2=20$, $R_1=5$, $R_2=10$ and $\alpha=0$.}
  \label{fig:Gauss1}
\end{figure}

\begin{figure}[htbp]
  \centering
  \includegraphics[scale=0.8]{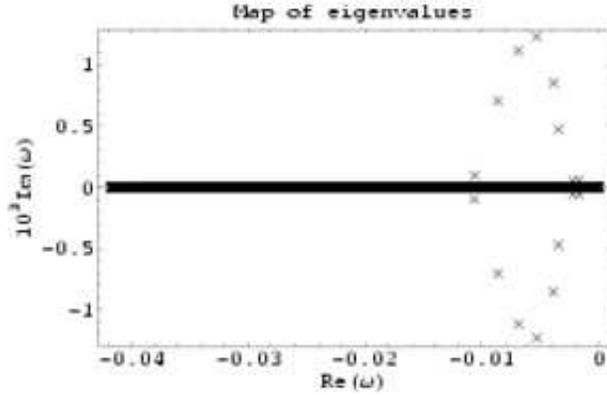}
  \caption{Same as Fig.~\ref{fig:Gauss1} but the inner wall has been removed, 
    i.e. $W_1=0$.}
  \label{fig:Gauss2}
\end{figure}

\begin{figure}[htbp]
  \centering
  \includegraphics[scale=0.8]{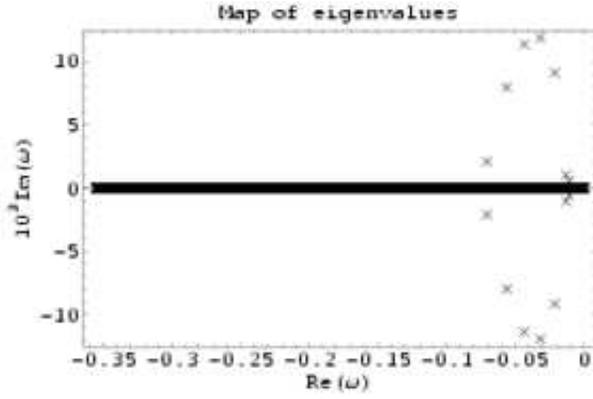}
  \caption{Same as Fig.~\ref{fig:Gauss1} but with a decreasing
    external magnetic field $\alpha=1$.}
  \label{fig:Gauss3}
\end{figure}

\begin{figure}[htbp]
  \centering
  \includegraphics[scale=0.8]{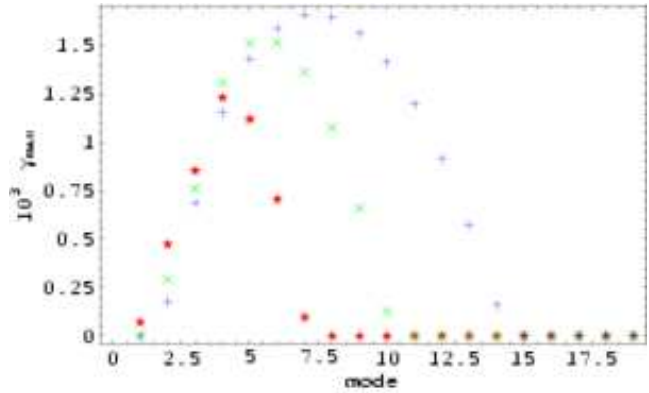}
  \caption{Comparison of the highest growth rates for different 
    spreads of the annular plasma column, $\sigma=1$ in red 'stars',
    $\sigma=0.6$ in green '$\times$' and $\sigma=0.4$ in blue '$+$'.}
  \label{fig:Gauss6}
\end{figure}

Indeed, taking boundaries such that $W_1=1$, $W_2=20$, $R_1=5$ and
$R_2=10$ and an uniform magnetic field $\alpha=0$, no charge on the
inner wall $Q=0$, the map of the eigenvalues in the complex plane is
given by Fig.~\ref{fig:Gauss1}.  Only the low azimuthal unstable modes
are excited, from $m=1$ to $m=7$, the fastest mode being $m=4$ with
$\gamma_{max} \approx 1.15 \times 10^{-3}$ which is of the same order
of magnitude as the real part, $Re(\omega)/m$.
Removing the inner wall has little effect on the diocotron spectrum,
compare Fig.~\ref{fig:Gauss1} and Fig.~\ref{fig:Gauss2}. Indeed, the
instability is initiated by the presence of a hole in the middle of
the plasma column. Placing or removing an inner wall cannot suppress
or generate the instability in this configuration. It can at most
affect the rate at which the instability grows.
Also, a decreasing magnetic field $\alpha=1$ does not affect the
results in a dramatic way, compare Fig.~\ref{fig:Gauss1} and
Fig.~\ref{fig:Gauss3}.
The number of excited modes is directly related to the radial spread
of the plasma column.  More precisely, the thiner the plasma ring, the
larger the number of unstable modes. In other words, decreasing
$\sigma$ will generate more unstable modes. In Fig.~\ref{fig:Gauss6},
we compare three Gaussian distributions with different spreads, namely
$\sigma = 1,~0.6,~0.4$.  The $\sigma=1$ profile has 7~unstable modes
whereas the $\sigma=0.6$ profile has 10~unstable modes and the
$\sigma=0.4$ profile 14.  Meanwhile, the highest growth rate is also
increasing significantly.

\begin{figure}[htbp]
  \centering
  \includegraphics[scale=0.8]{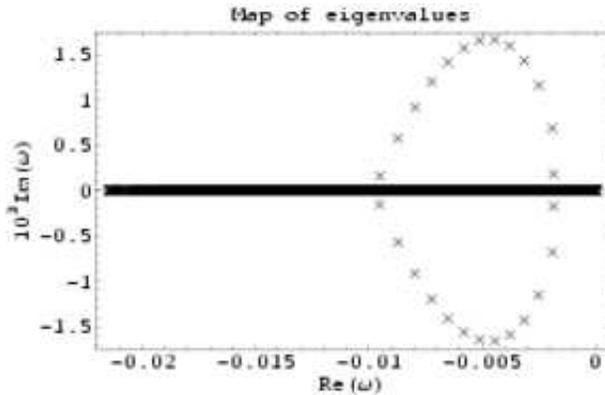}
  \caption{Same as Fig.~\ref{fig:Gauss1} but with a charge
    per unit length $Q=10$.}
  \label{fig:Gauss4}
\end{figure}

Finally, adding a charge to the inner wall does not affect drastically
the map of eigenfrequencies, see Fig.~\ref{fig:Gauss4}. More unstable
modes appear, nevertheless the maximum growth rates remain
approximately in the same order of magnitude.
Now that the algorithm has been check and employed for different lab
plasmas, we study in the next subsection the interesting case of an
electrospheric plasma.

\subsection{Electrosphere}
\label{subsec:Pulsar}

The electrospheric non-neutral plasma, as already proved in works
by~Krause-Polstorff and Michel~(\cite{1985MNRAS.213P..43K}) and
P\'etri et al.~(\cite{2002A&A...384..414P}), is confined by the
rotating magnetised neutron star. The most important feature is not
the density profile but the rotation speed, although both are related
in an unique manner, Eq.~(\ref{eq:DensiteCharge}).  Plasma shearing
motion between different magnetic surfaces leads to a kind of
Kelvin-Helmholtz instability, at least in the linear regime of the
instability (the eigenvalue problems look very similar).  (We always
keep a single domain decomposition such that $W_1=R_1$ and $W_2=R_2$).
Therefore, for electrospheric plasmas, we choose a rotation profile in
the plasma column that mimics the rotation curve obtained in the 3D
electrosphere. To study the influence of the profile, we took three
different analytical expressions for the radial dependence of $\Omega$
by mainly varying the gradient in differential shear as follows
\begin{equation}
  \label{eq:ProfilVit}
  \Omega(r) = 2 + \tanh[  \alpha \, ( r - r_0 ) ] \, e^{-\beta\,r^4}
\end{equation}
The values used are listed in Table~(\ref{tab:Vitesse}).
\begin{table}[htbp]
  \centering
  \begin{tabular}{cccc}
    \hline
    $\Omega$ & $\alpha$ & $\beta$ & $r_0$ \\
    \hline
    $\Omega_1$ & 3.0 & $5\times10^{-5}$ & 6.0 \\
    $\Omega_2$ & 1.0 & $5\times10^{-5}$ & 6.0 \\
    $\Omega_3$ & 0.3 & $5\times10^{-5}$ & 10.0 \\
    \hline
  \end{tabular}
  \caption{Parameters for the three rotation profiles used
    to mimics the azimuthal velocity of the plasma 
    in the electrospheric disk.}
  \label{tab:Vitesse}
\end{table}

\begin{figure}[htbp]
  \centering
  \includegraphics[scale=0.8]{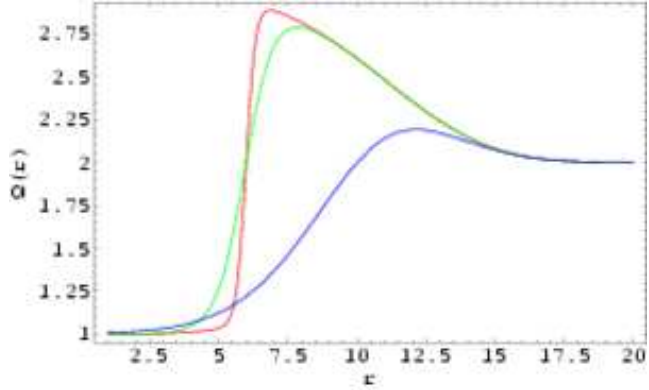}
  \caption{Three choices of differential rotation curves
    in the plasma column for the cylindrical pulsar electrosphere,
    $\Omega_1$ in red, $\Omega_2$ in green and $\Omega_3$ in blue.}
  \label{fig:OmegaElec}
\end{figure}

The angular velocity starts from corotation with the star $\Omega =
\Omega_* = 1$ (normalised to unity for convenience) followed by a
sharp increase around $r=6$ for $\Omega_{1,2}$ and a less pronounced
gradient around $r=10$ for $\Omega_3$.  Finally the rotation rate
asymptotes twice the neutron star rotation speed,
Fig.~\ref{fig:OmegaElec}.

\subsubsection{Uniform magnetic field}

\begin{figure}[htbp]
  \centering \includegraphics[scale=0.8]{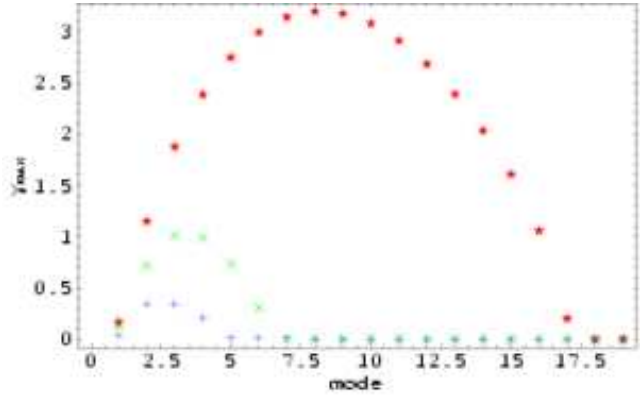}
  \caption{Fastest growth rate~$\gamma_{max}$ for each azimuthal mode $m$
    and the three rotation profile shown in Fig.~\ref{fig:OmegaElec}
    in an uniform external magnetic field, $\alpha=0$, $\Omega_1$ in
    red 'stars', $\Omega_2$ in green '$\times$' and $\Omega_3$ in blue
    '$+$'.}
  \label{fig:Elec2}
\end{figure}

First consider an uniform magnetic field, $\alpha=0$.  The maximum
growth rates for the three rotation curves for each mode~$m$ are shown
in Fig.~\ref{fig:Elec2}.  

The profile having the steepest gradient possesses the largest number
of excited unstable modes because it corresponds to the case where the
smallest scales appear, i.e. $m$ large, red stars in
Fig.~\ref{fig:Elec2}. The largest growth rate, for $m = 8$ its value
is $\gamma_{max} = 3.2$, is even larger than the maximum rotation
speed of the plasma column about $\Omega_{max} = 2.9$. The diocotron
instability operates on a very short timescale, comparable to the
period of the pulsar.

\begin{figure}[htbp]
  \centering \includegraphics[scale=0.8]{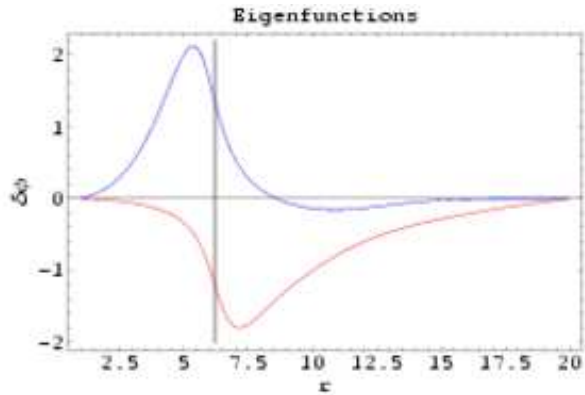}
  \caption{Real (red curve) and imaginary (blue curve) part 
    of the fastest growing eigenfunction for $\Omega_2$ in an uniform
    magnetic field. The vertical bar shows the location of the
    corotation radius.}
  \label{fig:Elec3}
\end{figure}

The second steepest profile possesses less unstable modes as we would
expect due to the fact that only larger scale structures can emerge
with this slope of the differential rotation, green~'$\times$' in
Fig.~\ref{fig:Elec2}.  The third smooth profile has only four unstable
modes, blue~'$+$' in Fig.~\ref{fig:Elec2}.
We give an example of eigenfunctions for the perturbed electric
potential in Fig.~\ref{fig:Elec3}. It corresponds to the fastest
unstable mode for the profile $\Omega_2$. The boundary conditions
impose $\phi(R_{1/2}) = 0$ as seen on the plot.  The corotation radius
$r_\mathrm{c}$ defined by
\begin{equation}
  \label{eq:CorotRad}
  \Omega(r_\mathrm{c}) = \frac{Re(\omega)}{m}
\end{equation}
is also shown on this plot, depicted by a vertical bar.

\subsubsection{Dipolar magnetic field}

\begin{figure}[htbp]
  \centering
  \includegraphics[scale=0.8]{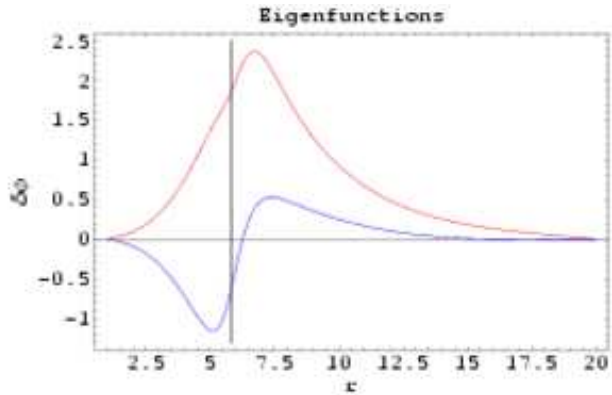}
  \caption{Real (red curve) and imaginary (blue curve) part 
    of the fastest growing eigenfunction for $\Omega_2$ in an dipolar
    magnetic field. The vertical bar shows the location of the
    corotation radius.}
  \label{fig:Elec5}
\end{figure}

Next consider a dipolar magnetic field, $\alpha=3$.  The maximum
growth rates for the three rotation curves for each mode~$m$ are shown
in Fig.~\ref{fig:Elec5}.

\begin{figure}[htbp]
  \centering
  \includegraphics[scale=0.8]{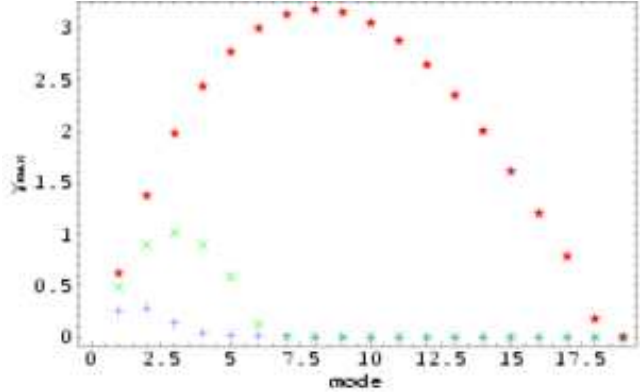}
  \caption{Fastest growth rate~$\gamma_{max}$ for each azimuthal mode $m$
    and the three rotation profile shown in Fig.~\ref{fig:OmegaElec}
    in a dipolar external magnetic field, $\alpha=3$, $\Omega_1$ in
    red 'stars', $\Omega_2$ in green '$\times$' and $\Omega_3$ in blue
    '$+$'.}
  \label{fig:Elec4}
\end{figure}

We can draw the same conclusions as in the previous case. The largest
number of excited unstable modes is observed for the steepest profile
$\Omega_3$.  Comparing Fig.~\ref{fig:Elec4} with Fig.~\ref{fig:Elec2},
the value of the maximum growth rate for each mode is not strongly
affected by the geometry of the magnetic field.
Here also, we give an example of eigenfunctions for the perturbed
electric potential in Fig.~\ref{fig:Elec5}. It corresponds again to
the fastest unstable mode for the profile $\Omega_2$. The boundary
conditions impose $\phi(R_{1/2}) = 0$ as seen on the plot.  Comparing
Fig.~\ref{fig:Elec5} with Fig.~\ref{fig:Elec3}, close to the inner
boundary, the real part of the eigenfunctions look very similar.
Nevertheless when approaching the outer boundary, the difference
becomes appreciable because of the strong departure from an uniform
magnetic field.

\section{CONCLUSION}
\label{sec:Conclusion}

We developed a pseudo-spectral Galerkin code to compute the
eigenspectra and eigenfunctions of the diocotron instability for
arbitrary magnetic field configurations and density profiles.  The
code is very efficient in computing the fastest growing modes. When
increasing the number of points, convergence is reached quickly, in
most cases, less than 200~points are needed.  Application to the
pulsar shows that the diocotron regime gives rise to instabilities
with growth rate comparable to the rotation period of the neutron
star. This confirms the results of our previous work (P\'etri et
al~\cite{2002A&A...387..520P}) and shows that the precise geometry, let
it be a flat disk or a column of plasma, does not affect drastically
the diocotron instability. The required ingredients are only
non-neutral plasmas and shear velocities.

In a forthcoming paper, we plan to investigate the full non-linear
development of the diocotron instability by means of 2D electrostatic
PIC simulations. We will also add an external source of charge feeding
the system and demonstrate that particle transport across the magnetic
field line is possible. Note that this has already been observed in
some experiments by Pasquini and Fajans~(\cite{2002AIPC..606..453P}).

Extension to plasma moving with relativistic speeds is also
envisageable, in particular the generalisation to the electromagnetic
non-neutral instability regime, the so-called magnetron instability.
Last but not least, the influence of finite temperature in the plasma
on the diocotron instability would require a kinetic treatment of the
stability via the Vlasov-Maxwell equation.

\begin{acknowledgements}
  I am grateful to Jean Heyvaerts for his helpful comments and
  suggestions. This work was supported by a grant from the G.I.F., the
  German-Israeli Foundation for Scientific Research and Development.
\end{acknowledgements}


\end{document}